\documentstyle[12pt,aasms4]{article}
\begin{document}

\title{Two Close--Separation Quasar--Quasar Pairs in the Large Bright Quasar Survey\altaffilmark{1}}
\author{Paul C. Hewett}
\affil{Institute of Astronomy, Madingley Road, Cambridge CB3 0HA, United Kingdom \\ Electronic mail: phewett@ast.cam.ac.uk}

\author{Craig B. Foltz}
\affil{Multiple Mirror Telescope Observatory, University of Arizona, Tucson, Arizona 85721 \\ Electronic mail: cfoltz@as.arizona.edu}

\author{Margaret E. Harding}
\affil{Institute of Astronomy, Madingley Road, Cambridge CB3 0HA, United Kingdom \\ Electronic mail: meh@ast.cam.ac.uk}

\author{Geraint F. Lewis\altaffilmark{2}}
\affil{Astronomy Program, Department of Earth and Space Sciences, SUNY
at Stony Brook, Stony Brook, New York 11794 \\ Electronic mail:
gfl@uvastro.phys.uvic.ca}

\altaffiltext{1}{Some of the observations reported here were obtained 
with the Multiple Mirror Telescope, a joint facility of the Smithsonian 
Institution and the University of Arizona}

\altaffiltext{2}{Present address: Astronomy Dept., University of
Washington, Box 351580, Seattle, WA 98195-1580, U.S.A.  and Dept.
of Physics and Astronomy, University of Victoria, PO Box 3055,
Victoria, BC V8W 3P6, Canada}

\begin{abstract}

We present photometric and spectroscopic observations of two close
quasar--quasar pairs found in the Large Bright Quasar Survey (LBQS)
(Hewett et al. 1995, AJ, 109, 1498). The two components of the
2153-2056 pair ($z=1.845$, $\Delta\theta=7\farcs8$, $B=17.9$ and
$21.3$) have the same redshifts within the relatively large
uncertainty, $\Delta v_{A-B} =-1100\pm1500{\rm\,km\,s^{-1}}$, of the
observations. The quasars are most likely spatially coincident although
the possibility that the pair is the result of gravitational lensing
cannot be ruled out. The two components of 1148+0055 ($z=1.879$,
$B=18.5$ and $z=1.409$, $B=21.1$, $\Delta\theta=3\farcs9$) have
disjoint redshifts and the pair has attracted some attention in the
context of gravitational lensing (e.g. Wampler 1997, ApJ, 476, L55)
following the independent discovery of the pair by Surdej and
collaborators. Four close, $\Delta\theta \le10\farcs0$, quasar--quasar
pairs have now been discovered in the LBQS and we discuss the
probability of identifying pairs with disjoint redshifts and of
locating spatially coincident pairs from the systematic investigation
of the well--defined quasar catalogue.

\end{abstract}

\section{INTRODUCTION}
The frequency and properties of close ($\lesssim 10\farcs0$) pairs of
quasars on the sky are of interest in the fields of gravitational
lensing, where the number of bona--fide lenses remains small
(\cite{keet96}), and the spatial clustering of quasars on scales
$\lesssim 100$\,kpc, where the number of such systems is also small
(\cite{croom96}). An investigation of the properties of close
companions to quasars in the Large Bright Quasar Survey (LBQS) (Hewett,
Foltz and Chaffee 1995) has produced four close, $<10\farcs0$, pairs of
quasars to date (Hewett, Harding and Webster 1992).  Investigations of
two pairs, 1429-0053 (\cite{hew89}) and 1009-0252 (\cite{hew94},
\cite{surdej94}), have already been published.  Here we present
spectroscopic and photometric observations of the two remaining pairs,
2153-2056 and 1148+0055. The close quasar pair 1148+0055 has
independently been discovered by \cite{surdej94} but no detailed
information has been published for either pair.

Section~\ref{obs} of the paper presents new spectroscopic and
photometric data for each of the latter two quasar pairs.  The
probability of identifying quasar pairs in the LBQS and comments on the
merits of gravitational lens models for the pairs are contained in
Section~\ref{disc}.

\section{OBSERVATIONS} \label{obs}

\subsection{Pair Identification}

The brighter images in both the 1148+0055 and 2153-2056 pairs were
identified as quasar candidates during the selection phase of the LBQS
and confirmed spectroscopically as quasars (\cite{hew91},
\cite{mor91}). Redshifts and magnitudes, $z=1.885$, $B_J=18.14$
(1148+0055) and $z=1.849$, $B_J=17.84$ (2153-2056) are listed in
\cite{hew95} with a note indicating that the presence of nearby
companion objects results in an overestimate of the quasar magnitudes
and an error of up to $3\farcs0$ in the celestial coordinates.

As part of an investigation into the frequency of strong gravitational
lenses in the LBQS all the quasars were inspected visually to provide a
census of companion objects within a radius of $10\farcs0$. Film copies
of the UKST $B_J$ direct plates scanned for the LBQS project, or, if
available, better quality $B_J$ survey plates, were used and companions
to $B_J=21.5-22$, depending on the quality and depth of the $B_J$
plate, were identified. In practice, the physical extent of the high
density images of the relatively bright LBQS quasars on the $B_J$
survey plates precludes the detection of faint companions closer than
$\sim 3\farcs0$. 1148+0055 was noted as possessing a faint stellar
companion with a separation of $\sim 4\farcs0$, and 2153-2056 a
companion with a separation of $\sim 7\farcs5$.

\subsection{2153-2056: Spectroscopy}

Follow--up spectroscopy of both components of the 2153-2056 pair was
obtained at the Anglo-Australian Telescope on 1989 August 29.
Observations were made with the RGO Spectrograph and 25cm Camera,
employing a $600\,{\rm g}\,{\rm mm^{-1}}$ grating, and the IPCS as the detector. Three
exposures, $2500\,$s and $2 \times 3600\,$s, were obtained. The
slit--width was $1\farcs8$ and the slit was oriented at
PA=152$^{\circ}$ across both members of the pair.
The spectrograph configuration, slit--width and detector combination
gave a wavelength resolution of $\sim 8$\AA \ and a usable wavelength
coverage of $\sim 3800-5800$\AA. The spatial scale along the slit was
$0\farcs4$ per pixel. Atmospheric transmission and the seeing, $\sim
1\farcs0$, were both good throughout the period of the observations.

Data reduction was performed using standard procedures available in the
Starlink {\tt FIGARO} spectroscopic reduction package. The primary and
secondary spectra are well separated on the detector and no particular
difficulties were encountered during the reduction. A flatfield was
obtained from an exposure of a tungsten lamp and wavelength calibration
was based on exposures of a CuAr arc-lamp before and after the
observations of 2153-2056. A standard atmospheric extinction correction
was applied to the spectra, but, given the low signal--to--noise ratio
of the companion spectra, no attempt to perform absolute or relative
fluxing was made. Following the extraction of the one--dimensional,
extinction corrected, spectra from the individual frames the spectra
were summed to produce the spectra shown in Figure 1.

Compared to the discovery spectrum (\cite{mor91}) the AAT spectrum of
the LBQS quasar has more limited wavelength coverage but a higher
signal--to--noise ratio. The two spectra are consistent and the quasar
is an unremarkable representative of the class of
optically--selected quasars. The redshift derived from the AAT spectrum
via cross--correlation with a composite spectrum (\cite {francis91}) is
$z=1.845\pm 0.003$, compared to $z=1.849\pm 0.005$ reported in
\cite{hew95} based on the discovery spectrum.

The spectrum of the companion object (B) has poor signal--to--noise
ratio but the presence of broad emission features at $\sim 4400$\AA
\ and $5400$\AA \ makes its identification as a quasar with $z\sim
1.85$ unambiguous. Scaling the spectrum of the LBQS quasar (A) to the
same mean count level over the wavelength range $4250-5500$\AA
\ confirms the fainter object has a very similar overall continuum
shape and comparable emission line properties. A cross--correlation of
the two spectra produces a velocity difference $\Delta v_{A-B}
=-1100\pm1500{\rm\,km\,s^{-1}}$ with the spectrum of B redshifted
relative to that of A. The spectra are thus consistent with both
components lying at the same redshift but with a rather large
uncertainty.

The most obvious difference between the two spectra is the extended
depression in the continuum of the B component centred at $\sim
4230$\AA. The feature is visible in the individual sky--subtracted
two--dimensional frames and is certainly real. The characteristics of
the absorption are typical of a \ion{C}{4} $\lambda1550$ broad
absorption line (BAL) trough with a peak ejection velocity of $\sim
18,000{\rm\,km\,s^{-1}}$.  This interpretation is reinforced by the
apparent turn--down at $\sim 3950$\AA \ in the spectrum of B which is
consistent with the presence of a \ion{Si}{4} $\lambda1400$ BAL
trough.

The spectrum of the A component shows weak absorption features at
$4308.3$\AA \ and $4316.2$\AA, the ratio of the wavelengths is $1.00183$,
close to the $1.00166$ for the \ion{C}{4} $\lambda\lambda 1548.2,
1550.8$ doublet. The probable identification is with a \ion{C}{4}
doublet at $z=1.783$ and rest equivalent width $\sim 0.7$\AA.

\subsection{2153-2056: Photometry}

Optical $B$, $V$, $R$ and $I$ magnitudes were obtained using the 2.5m
Isaac Newton Telescope (INT) on La Palma on 1989 September 5 ($R$ and
$I$) and 1989 September 8 ($B$ and $V$). An RCA CCD with $30\mu$m pixels,
giving an image scale of $0\farcs74$ per pixel, was employed as the
detector. KPNO filters were used with observations of standards taken
from \cite{chris85}. Bias frames were obtained during both nights,
while flatfields were derived from twilight sky exposures. Observations
of standards at the beginning and end of the nights, and at intervals
throughout, were obtained to allow determination of the extinction and
color transformations. Observations of 2153-2056 were obtained at
airmass, $1.5-1.6$. Transparency was good on both nights and the seeing
was excellent, although the full width half maxima (FWHM) of images
measured from CCD frames of the quasar pair was limited by the large
pixels to $\sim 1\farcs4$. Exposure times were $3 \times 200\,$s ($B$),
$3 \times 200\,$s ($V$), $8 \times 300\,$s ($R$) and $2 \times 300\,$s
($I$).

\setcounter{footnote}{0}
Data reduction was performed using standard procedures within {\tt
IRAF}~\footnote{{\tt IRAF} is distributed by the National Optical
Astronomy Observatories, which are operated by the Association for
Research in Astronomy, Inc. under contract with the National Science
Foundation} Magnitudes were derived using M.J. Irwin's {\tt IMAGES}
software (\cite{irwin82}, \cite{irwin85}). Extinction, zero--points and
color transformations were well determined. The errors of some of the
standard magnitudes from \cite{chris85} are substantial but the
residuals of the standards about the derived fits are consistent with a
contribution to the rms scatter from the CCD observations of $\lesssim
0.02\,$mag. The resulting Johnson $B$ and $V$ magnitudes and
Kron--Cousins $R$ and $I$ magnitudes are given in Table~1. The quoted
errors include uncertainties in the zero--point determination and color
transformations.

The B1950.0 coordinates of the A component derived by matching
stars visible on the CCD frames to the corresponding images in the APM
scan of the UKST plate are $21^h 53^m 06\fs57$ $-20\arcdeg 56\arcmin
03\farcs1$, approximately $0\farcs9$ south--east of the position quoted
in \cite{hew95}. This difference is consistent with the bias induced by
the presence of the faint companion, which is merged with the brighter
component in the APM scan. The separation of the two components derived
from the $V$ and $R$ frames is $\Delta\theta = 7\farcs8\pm0.1$ and the
position angle of the fainter component relative to the brighter
component is PA$=331^{\circ}\pm2$. Figure 2 shows a $134\farcs0 \times 134\farcs0$ region centred on the A component of the pair from the coadded $R$--band CCD frame. 

The CCD frames have $1\sigma$ sky fluctuations within a $4\farcs4$
diameter aperture of $24.6$, $24.3$ and $25.1\,$mag for the $B$, $V$
and $R$ frames respectively. There is no evidence for the presence of
any additional images, apart from the B component of the pair, within a
$15\farcs0$ radius of component A. However, the very large pixels
preclude the detection of faint images close to component A via point
spread function subtraction techniques and it is not possible to rule
out the presence of an image fainter than $R \sim 22.0$ within $1\farcs5$
of component A.

The number counts in the field turn over at $R\sim 24.0$ and the image
catalogue appears to be largely complete to $R=23.8$. There are
293 images with magnitudes $21.0 \le R \le 23.8$ in an area of
$0.0067\,{\rm deg}^2$. Reliable image classification is not possible
for the fainter objects but the CCD frame, which is at
Galactic coordinates $l,b=32,-50$, contains significantly more stars
than a typical high--latitude field. The $R$--band galaxy counts of
\cite{met95}, predict 243 galaxies, magnitudes $21.0 \le R \le 23.8$ in
an area of $0.0067\,{\rm deg}^2$. Allowing for the presence of
stars in the image catalogue the galaxy number counts are in excellent
agreement and there is no evidence for a significant excess.  The
spatial distribution of objects over the frame is uniform on large
scales with fluctuations evident on scales of $30\farcs0-60\farcs0$.
The quasar pair lies in a region that is somewhat underdense relative
to the mean. 

Infrared $J$, $H$ and $K_{n}$ magnitudes were obtained using IRIS at
the AAT on 1992 October 13. A $128 \times 128$ Rockwell HgCdTe array
with $60\mu$m pixels, producing an image scale of $0\farcs6$ per pixel,
was employed as the detector. 2153-2056 was observed in $H$ and $K_{n}$
using a five--point dither with an offset of $10\farcs0$ in right
ascension and declination between positions. Exposure times were
$125\,$s per position, giving total exposure times of $600\,$s in both
$H$ and $K_{n}$. A two--position, $25\farcs0$ separation, $125\,$s
exposure, strategy was adopted for the $J$ observations to give a total
exposure time of $250\,$s per target. The $K_{n}$ filter has a narrower
FWHM, extending over the range $2.0-2.3\mu$, and lower effective
wavelength in comparison to the conventional $K$ filter. A transmission
curve for the $K_{n}$ filter can be found in the IRIS User Guide
(\cite{iris}).

All frames were first linearized and dark subtracted using a standard
procedure (\cite{iris}). Subsequent data reduction was
accomplished using {\tt IRAF} routines. A normalized flatfield, derived
separately for each filter, from exposures of an illuminated
dome--screen at the start of the night, was divided into all the object
exposures. Sky--frames were then generated by combining groups of
dithered object exposures. A scaled version of the sky--frame was
subtracted from each object frame and the resulting sky--subtracted object
frames were combined using integer pixel shifts in the spatial
registration procedure and a mask to ensure the small number of bad
pixels in the array did not contribute to the final summed image. SAAO
standards (\cite{cart95}) were observed throughout the night to provide
zero--points (on the SAAO system) and extinction values. Magnitudes
were derived using {\tt IMAGES} software and are given in Table~1.  The
quoted errors include uncertainties in the zero--point determination.

The fainter B component of the pair is just visible in all three bands
but the flux excess in a $3\farcs6$ diameter aperture is only
$2-3\sigma$ of the sky background fluctuations. Relative photometry of
both members of the pair using a $3\farcs6$ aperture gives magnitudes
for the companion of:  $J=20.5\pm1.0$, $H=19.2\pm0.5$ and
$K_{n}=18.3\pm0.4$. A composite frame, made from averaging the $J$, $H$
and $K_{n}$ frames, produces a relative magnitude difference between
the components of $\Delta m_{B-A} = 3.3\pm0.3$. Thus, within the very
poor constraints, the optical to infrared colors of both components
are similar.

No other images are visible in the infrared frames within a radius of
$30\farcs0$ of the the A component.

\subsection{1148+0055: Spectroscopy}

Follow--up spectroscopy of both components of the 1148+0055 pair was
obtained at the Multiple Mirror Telescope (MMT) on 1990 June 17. The Red
Channel of the MMT Spectrograph was used with  a $150\,{\rm g}\,{\rm mm^{-1}}$ grating and
a thinned TI $800\times 800$ CCD as the detector. One exposure, of
$1200\,$s duration was obtained. The slit--width was $1\farcs25$ and
the slit was oriented at PA=105$^{\circ}$ across both primary and
secondary members of the pair. The spectrograph configuration,
slit--width and detector combination gave a wavelength resolution of
$\sim 20$\AA \ and a usable wavelength coverage of $\sim
3200-7750$\AA.  The spatial scale along the slit was $0\farcs6$ per
pixel. Atmospheric transmission and the seeing, $\sim 1\farcs0$, were
both good throughout the period of the observations.

Data reduction was performed using standard {\tt IRAF} procedures. The
primary and secondary spectra were well separated on the detector and
no particular difficulties were encountered during the reduction. A
flatfield was obtained from an exposure of a quartz lamp and wavelength
calibration was based on exposures of a HeNeAr arc-lamp taken after the
observation of 1148+0055. A standard atmospheric extinction correction
was applied to the spectra. The observations were flux calibrated using
observations of standard stars but no attempt to determine absolute
fluxes was made.

The observations were taken at an airmass of 1.8, the slit was not
aligned at the parallactic angle and the MMT's intensified television
camera is sensitive primarily to red wavelengths. Thus, notwithstanding
the flux calibration procedure, the quasar spectra show a significant
loss of light shortward of about $4500$\AA \ due to the effects of
atmospheric dispersion. To correct for the systematic loss of flux the
spectra were first divided by the composite spectrum of
\cite{francis91}, the continuum flux distribution of which is similar
to that of the LBQS quasar 1148+0055 (\cite{hew91}).  A low--order
polynomial was fitted to the result and divided into the original
spectra to produce the spectra shown in Figure 3. The shape of the
corrected spectrum of 1148+0055 and the original discovery spectrum are
in good agreement and the relative fluxes should be good to $\sim 20\%$
or better.

The redshift of the LBQS quasar, component A, derived via
cross--correlation with the LBQS composite spectrum is $z=1.879\pm
0.003$, compared to $z=1.885\pm 0.005$ reported in \cite{hew95}, based
on the discovery spectrum. The emission line and continuum properties
are not at all unusual for an optically--selected quasar.

The spectrum of the B component shows strong broad emission at $\sim
3730$, $4580$ and $6750$\AA \ and, combined with the presence of an
extended continuum, identification as a quasar with $z\sim 1.4$
exhibiting \ion{C}{4} $\lambda1550$, \ion{C}{3}] $\lambda1909$ and
\ion{Mg}{2} $\lambda2798$ emission lines is unambiguous. The
cross--correlation redshift is $z=1.409\pm 0.003$.  Within the
constraints of the relatively poor signal--to--noise ratio there is
nothing remarkable about the spectrum.

The spectrum of the A component was searched for absorption arising in,
or associated with, the lower redshift B component. At a redshift of
$z=1.409$, \ion{C}{4} $\lambda\lambda1548,1550$ and \ion{Mg}{2}
$\lambda\lambda 2796,2803$ doublets at about $3734$\AA \ and $6745$\AA{} fall
within the wavelength coverage of the spectrum of the A component.  No
significant features were seen. Given the relatively low spectral
resolution and signal--to--noise ratio of the spectrum, the rest
equivalent width limit on an unresolved absorption line is a relatively
large $3$\AA{} and a higher quality spectrum is necessary for
absorption line detections or to establish interesting upper limits.

\subsection{1148+0055: Photometry}

\setcounter{footnote}{0}
Optical $B$, $V$ and $R$ magnitudes were obtained using the 1m Jacobus
Kapteyn Telescope (JKT) on La Palma on 1992 April 1 ($B$ and $V$) and
1994 May 2 ($V$ and $R$). In 1992, a GEC CCD with $22.0\mu$m pixels,
giving an image scale of $0\farcs30$ per pixel, was employed as the
detector\footnote{Note that the CCD employed on this night was
incorrectly identified as an EEV $1242 \times 1152$ pixel CCD in
\cite{hew94}. Although the correct image scale was employed to
calculate the separation of the images in the 1009-0252 system, the
relative separations of component B from A listed in Table 1 should
read (-0.61,-1.41) in agreement with the caption to Figure 1 of
\cite{hew94}. This inconsistency has previously been noted by
\cite{keet96}}.  KPNO filters were used with observations of standards
taken from \cite{land83}. In 1994, an EEV CCD with $22.5\mu$m pixels,
giving an image scale of $0\farcs31$ per pixel was employed as the
detector.  Harris filters were used with observations of standards
taken from \cite{land92}. Bias frames were obtained during both nights
and flatfields were derived from twilight sky exposures. Observations
of standards at the beginning and end of the nights, and at intervals
throughout, were obtained to allow determination of the extinction and
color transformations. Observations of 1148+0055 were obtained at low
airmass, $\lesssim 1.2$, transparency was good on both nights although
in 1992 there was a small systematic variation in the vertical
extinction of $0.007\pm0.001\,$mag per hour. The full width half maxima
of images measured from CCD frames of the quasar pair were $\lesssim
1\farcs0$. Exposure times were $2 \times 500\,$s ($V$) and $2 \times
1000\,$s ($B$) in 1992 and $2 \times 500\,$s ($V$) and $2 \times
500\,$s ($R$) in 1994.

Data reduction was performed using standard procedures within {\tt
IRAF}. Magnitudes were derived using M.J. Irwin's {\tt IMAGES}
software. Extinction, zero--points and color transformations were well
determined with residuals of the standards about the derived fits
typically $\lesssim 0.02\,$mag. The resulting Johnson $B$ and $V$
magnitudes and Kron--Cousins $R$ magnitude are given in Table~1. The
quoted errors include uncertainties in the zero--point determination
and color transformations. The $R$ magnitudes of the components are in
good agreement with those published by \cite{surdej94}.

The B1950.0 coordinates of the A component derived by matching stars
visible on the CCD frames to the corresponding images in the APM scan
of the UKST plate are within $0\farcs1$ of the position quoted in
\cite{hew95}, i.e., $11^h 48^m 41\fs56$ $+00\arcdeg 55\arcmin
07\farcs8$. The presence of the faint companion, which is merged with
the brighter component in the APM scan, had essentially no effect on
the centroid position of the brighter component. The separation of the
two components derived from the $B$, $V$ and $R$ frames is
$\Delta\theta = 3\farcs9\pm0.1$ and the position angle of the fainter
component relative to the brighter component is PA$=111^{\circ}\pm2$.
The separation of the components is in excellent agreement with that
published by \cite{surdej94}. 

The CCD frames are not particularly deep, with $1\sigma$ sky
fluctuations within a $3\farcs6$ diameter aperture of $24.0$, $23.5$
and $23.2\,$mag for the $B$, $V$ and $R$ frames respectively. There are
a number of faint galaxies, $V \sim 22.5$, $R\sim 21$, in the annulus
$20\farcs0 - 60\farcs0$ from the pair but there is no significant
enhancement of images or otherwise unusual feature of the immediate,
$\sim 1\farcm0$, neighborhood. The closest object to the pair, $\sim
18^{\prime\prime}$ distant and PA$\sim 304^{\circ}$ from component A,
is compact with magnitudes, $R=20.9\pm0.2$, $V=22.7\pm0.3$. Figure 4
shows a $134\farcs0 \times 134\farcs0$ region centred on the A
component of the pair from the coadded $R$--band CCD frame.

Infrared $J$, $H$ and $K$ magnitudes were obtained using UKIRT on 1994
February 24. A $58 \times 62$ Santa Barbara Research Corporation InSb
array with $15\mu$m pixels, producing an image scale of $0\farcs6$ per
pixel, was employed as the detector. 1148+0055 was observed in each
band using a 5--point dither with an offset of $8\farcs0$ in right
ascension and declination between positions. Transparency was good and
the seeing was excellent, although the FWHM of images measured from CCD
frames of the quasar pair was limited by the size of the detector
pixels to $\sim 1\farcs1$.  Exposure times were $120\,$s, $80\,$s and
$90\,$s per position, giving total exposure times of $600\,$s, $400\,$s
and $450\,$s in $J$, $H$ and $K$ respectively.  Data reduction was
accomplished using {\tt IRAF} routines and involved subtraction of a
dark frame from each exposure, followed by combination of the dithered
object exposures to produce a flatfield frame. A normalized version of
the flatfield was then divided into each object frame. The resulting
object frames were combined using integer pixel shifts in the spatial
registration procedure and a mask to ensure the small number of bad
pixels were excluded from the final summed image.  Standards from the
UKIRT bright standard list were observed throughout the night to
provide zero--points (on the UKIRT system) and extinction values.
Magnitudes for the bright member of the pair were derived using {\tt
IMAGES} and are given in Table~1. The quoted errors include
uncertainties in the zero--point determination. 

The fainter B component of the pair is present in all three bands at
$5-7\sigma$ of the sky background fluctuations. Relative photometry of
both members of the pair using a $3\farcs6$ diameter aperture was used
to derive the magnitudes for the companion given in Table~1. The
overall optical to infrared colors of both components are similar.  The
infrared colors of the B component appear to be significantly redder
than those of the A component although the errors are substantial.

There is one other object visible within the small $22\farcs0 \times
20\farcs0$ area of the $J$, $H$ and $K$ frames. The object is
$5\farcs0\pm0.2$ from A at PA$=276^{\circ}\pm2$, placing it almost
opposite component B. The object is well detected in the $J$ and $K$
frames but barely present in the $H$ frame. Magnitudes, derived using
differential photometry in a $3\farcs6$ diameter aperture relative to
component A, are $J=20.4\pm0.2$, $H=21.2\pm1.0$ and $K=18.4\pm0.2$.
Inspection of the $R$--band CCD frame reveals a suggestion of something
coincident with the infrared detection but far too faint to perform any
form of photometry.

Taking a $K$--band detection limit in the UKIRT infrared frames of
$K=19.0$ the predicted surface density of galaxies within a radius of
$\Delta\theta=10\farcs0$ from the LBQS component of 1148+0055 is $\sim
0.4$, where the galaxy surface density is taken from \cite{mcleod95a}.
The detection of one such object is thus not unexpected. The
photometric data is poor but the $K$ magnitude, $J-H$ color, and the
limit to the optical to infrared color are consistent with
identification as a luminous, $L^*$, early--type galaxy at redshift
$z\sim 1$ (see magnitude and color predictions in \cite{mcleod95b} and
\cite{song94}). However, much higher quality optical and infrared
photometry, or a spectroscopic redshift, is required to classify the
object.

\section{DISCUSSION} \label{disc}

The discovery of close quasar--quasar pairs with discordant redshifts
has stimulated considerable interest. Specifically, the quasar--quasar
pair 1148+0055, along with the 1009-0252 pair (\cite{hew94}), has been
discussed in the context of weak gravitational lensing (\cite{burb97},
\cite{wamp97}), although it has not been appreciated that both pairs
were discovered among a well defined sample of quasars, i.e., the
LBQS.  The selection effects that determine the probability of locating
close, $\lesssim 10\farcs0$, pairs of objects in the LBQS are not
simple and the spectroscopic follow--up of the large number of faint
objects around the 1055 LBQS quasars is a non--trivial undertaking.
Consequently the definitive analysis of the frequency and properties of
quasar--quasar and quasar--galaxy pairs in the LBQS is not yet
available. However, it is possible to perform a reliable {\it a priori}
calculation of the number of companion quasars expected by chance to be
found within an annulus, $3\farcs0 \le \theta \le 10\farcs0$, about all
of the 1055 LBQS quasars. The search for companion quasars performed on
the entire LBQS catalogue is sensitive to companions with magnitudes
$B_J \le 21.5$, redshifts $0.2 \le z \le 3.4$, and separations
$3\farcs0 \le \theta \le 10\farcs0$. The surface density of such
quasars to $B_J=21.5$ is $\sim 70\,{\rm deg^{-2}}$ (\cite{hart90}) and
the total area surveyed is $0.0233\,{\rm deg^2}$, giving a predicted
number of $1.6$ unrelated quasar--quasar pairs within the LBQS. The
follow--up spectroscopy of viable candidates for companion quasars is
almost complete; the efficiency of the identifications within the
specified magnitude and separation limits is close to $100\%$ and it is
unlikely that further pairs will be identified. Thus, the detection of
two quasar--quasar pairs in the survey is not unexpected. This
conclusion is at variance with that reached by \cite{burb97}. The
difference is attributable to a combination of factors, including the
values taken by \cite{burb97} for the number of quasars surveyed (648
c.f. 1055), the smaller maximum separation limit ($\theta < 5\farcs0$
c.f. $3\farcs0 \le \theta \le 10\farcs0$) and the brighter quasar
magnitude limits ($m\le 19.3$ and $m\le20.7$ c.f. $B_J \le 21.5$). It
is true that both the quasar--quasar pairs discovered in the LBQS have
separations $\theta < 5\farcs0$ and that one of the companion quasars
is relatively bright.  However, the frequency of quasar--quasar pairs
found in the LBQS is consistent with the {\it a priori} calculation
based on the well--defined survey parameters.

The rather high probability of finding close pairs of quasars with
discordant redshifts in the LBQS catalogue by chance does not mean that
gravitational lensing of quasars by mass associated with foreground
quasars is unimportant and \cite{wamp97} reviews the lensing
hypothesis. If the unidentified object detected in the infrared frames
is a high--redshift $z\sim 1$, massive, early--type galaxy, it's
effect, combined with that of the mass associated with the
low--redshift quasar, could produce a magnification of the LBQS
quasar. However, the magnification is not expected to be very large,
$\lesssim 0.5\,$mag, as there is no evidence for the presence of
multiple images and the cross--section for an isothermal deflector
producing a magnification of even a factor two is very small.

The 2153-2056 pair ($z=1.85$, $\Delta\theta=7\farcs8$, $B=17.9$ and
$21.3$) has a number of similarities to the 1429-0053 pair ($z=2.08$,
$\Delta\theta=5\farcs1$, $B=17.7$ and $20.8$, \cite{hew89}) also found
in the LBQS catalogue. Such pairs may arise as the result of the
spatial clustering of quasars (\cite{croom96}), or, due to
strong gravitational lensing by a massive, $>10^{12}\,M_{\sun}$,
intervening object.

\setcounter{footnote}{0}
At $z=2.0$, $1\farcs0 \simeq 5\,{\rm h}^{-1}\,{\rm kpc}$, for
cosmologies with $\Omega =0.3-1$ and zero cosmological
constant~\footnote{{\rm h} is the Hubble constant in units of
$100\,{\rm kms^{-1}}{\rm Mpc^{-1}}$}. Little is known about the
clustering properties of quasars on the small spatial scales, $\lesssim
50\,{\rm h}^{-1}\,{\rm kpc}$, corresponding to the $\le 10\farcs0$
separations of the LBQS quasar--quasar pair data.  Extrapolating the
observationally determined properties of quasar clustering on
megaparsec scales, where the spatial two--point correlation function at
separation $r\,{\rm h}^{-1}\,{\rm Mpc}$ is $\xi \sim (r/r_0)^{-1.8}$,
$r_0=6\,{\rm h}^{-1}\,{\rm Mpc}$ (\cite{croom96}), produces an excess
of four orders of magnitude in the number of quasars expected within
$50\,{\rm h}^{-1}\,{\rm kpc}$ of a quasar, relative to that expected
for an unclustered population. The faint apparent magnitude limit
probed, $B_J\sim 21.5$, means companion objects well down the
luminosity function can be detected ($M_{B_J} \sim -23.8$ at $z=2.0$
for h=$0.5$ and $\Omega=1$) and the space density is relatively high,
$\sim 10^{-6}$ quasars per Mpc$^3$. However, as stressed by Djorgovski
(1991) even the high space density coupled with the enhanced
probability of finding a quasar within $\le 100\,{\rm h}^{-1}\,{\rm
kpc}$ falls several orders of magnitude short of explaining the number
of common redshift close pairs found in various quasar surveys. The
statistics for the LBQS survey confirm the conclusions of Djorgovski
(1991), with only $\sim 0.01$ quasar pairs predicted compared to the
two observed. As noted in that study, there is no reason why a simple
extrapolation of the clustering behavior evident at large separations
in the linear regime should be valid at close separations where several
additional physical processes may be important, including the
possibility of close encounters triggering enhanced fueling of active
galactic nuclei. The very large surveys, $\sim 20,000$ quasars, planned
for the Anglo-Australian Telescope's 2dF multi--fiber instrument should
place the empirically determined behavior of quasar clustering on small
scales $\lesssim 1 {\rm h}^{-1}\,{\rm Mpc}$ on a much firmer footing,
thereby allowing a much improved assessment of whether the number of
detected quasar--quasar pairs at very small scales, $\lesssim 100 {\rm
h}^{-1}\,{\rm kpc}$, is consistent with the spatial clustering of
quasars.
 
Under the gravitational lensing hypothesis only the separation of the
2153-2056 quasar pair and their relative fluxes~\footnote{Any possible
microlensing induced magnification due to the action of individual
stars in a lensing galaxy has been neglected}, provide constraints on
the model. Assuming the deflector can be modelled as a singular
isothermal sphere (\cite{schneid92}) and lies at a redshift $z=1.0$,
the gravitational lensing solution is determined exactly. The velocity
dispersion of the deflector is ${\sigma_v \sim 690}{\rm km~s^{-1}}$,
equivalent to that of an Abell Richness=1 cluster (\cite{zab90}). The
centroid of the deflector lies between the two quasars, $\sim 7\farcs4$
away from the brighter system. The ability to detect a relatively poor
Abell cluster at $z \gtrsim 1$ with images of the depth of the INT
$R$--band frame is poor (\cite{post96}) and the photometric data is
consistent with the presence of such a cluster but offers no positive
evidence to support the lensing hypothesis. The most probable redshift
for a deflector is $z_d \sim 0.6$, significantly lower than the
$z_d=1.0$ employed above, reducing the velocity dispersion of the
deflector to $\sim 520{\rm km~s^{-1}}$, but the lack of any evidence
for an overdensity of galaxies arising from the presence of a cluster
at $z\sim 0.6$ mitigates against such a geometry and offers no evidence
in favor of the lensing hypothesis.
 
Macrolensed BAL quasars can provide a valuable insight on the nature
of the absorbing material. In the case of 2153-2056, the path
separation to the two images is $\sim 5\times10^{15}$~cm for a BAL
region $200\,$pc from the quasar core.  The time--delay between the
images from the model with a deflector at $z_d=1.0$ is $\sim 16 {\rm h^{-1}}$ years. This value
is reasonably insensitive to cosmology and such a time--delay, coupled
with intrinsic variability of the quasar source, could conceivably
account for the observed spectral differences between the images.

Considering individual systems it is difficult to rule out the lensing
hypothesis completely, however, under the lensing hypothesis it is
predicted that lensed image pairs with similar flux ratios should be
more prevalent than image pairs with more disparate fluxes
(\cite{koch95}). The selection of the 1429-0053 and 2153-2056 pairs
with their large brightness ratios, without the detection in the LBQS
of a larger number of quasar pairs with more similar fluxes, argues
against the lensing hypothesis for the origin of the pairs.
  
\acknowledgements

The LBQS is supported by National Science Foundation Grant Nos.
AST-90-01181 and 93-20715. The LBQS would not have been possible
without the active support of of the United Kingdom Schmidt Telescope
Unit and the staff of the Automated Plate Measuring facility. We are
grateful to the referee, Chris Kochanek, for highlighting the
difference between the predicted and observed flux ratios of the quasar
pairs under the lensing hypothesis.  The authors acknowledge the data
analysis facilities provided by the Starlink Project which is run by
CCLRC on behalf of PPARC. A NATO Collaborative Research Grant aids
research on gravitational lenses and quasar surveys at the Institute of
Astronomy. The INT and JKT are operated on the island of La Palma on
behalf of PPARC and NWO at the Observatorio del Roque de los Muchachos
of the Instituto de Astrofisica de Canarias. Telescope operators at the
AAT, MMT and La Palma provided valuable support and we would also like
to thank Rachel Webster who helped obtain the spectra of 2153-2056 at
the AAT.

\newpage
This page left blank for Table 1.
\newpage
\center{\bf FIGURE CAPTIONS}
\bigskip

\figcaption{Top panel: AAT + RGO Spectrograph spectra of LBQS
2153-2056A,B.  No flux calibration has been applied and for the purpose
of this figure, the data for B have been multiplied by a factor of 15
and smoothed with a 3-pixel wide running boxcar.  Note the presence of
a clear \ion{C}{4} broad absorption trough in the spectrum of component
B centered at about 4250\AA. Bottom panel:  Uncalibrated spectra
binned into 15\AA{} pixels and overplotted.  Note that, longward of
about $4500$\AA, the two spectra are similar to one another; shortward
of $4500$\AA, the spectrum of B is affected by strong broad
absorption.}

\figcaption{$R$--band CCD image of a $134\farcs0 \times 134\farcs0$ region
centred on the A component of the quasar pair 2153-2056. North is to
the top with east to the left. The quasar components are indicated by
the labels ``A'' and ``B''.}

\figcaption{MMT Red Channel spectra of LBQS 1148+0055A,B.  For the purpose
of this figure, the data for B have been multiplied by a factor of 3
and smoothed with a 3-pixel wide running boxcar.  The continuum shapes of 
the spectra have been adjusted to that of the LBQS composite spectrum
(see text for details).}

\figcaption{$R$--band CCD image of a $134\farcs0 \times 134\farcs0$ region
centred on the A component of the quasar pair 1148+0055. North is to
the top with east to the left. The quasar components are indicated by
the labels ``A'' and ``B''.}

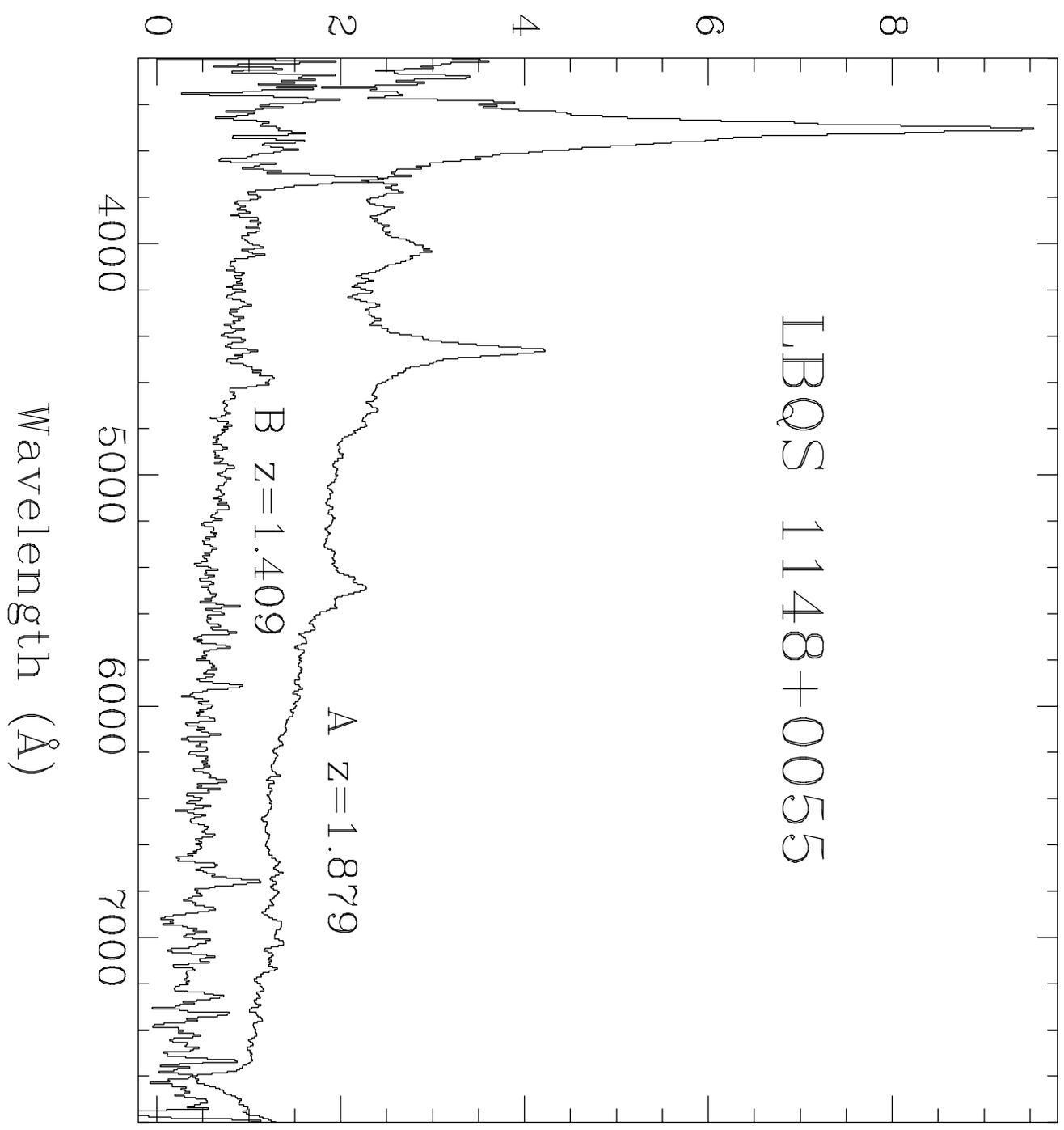
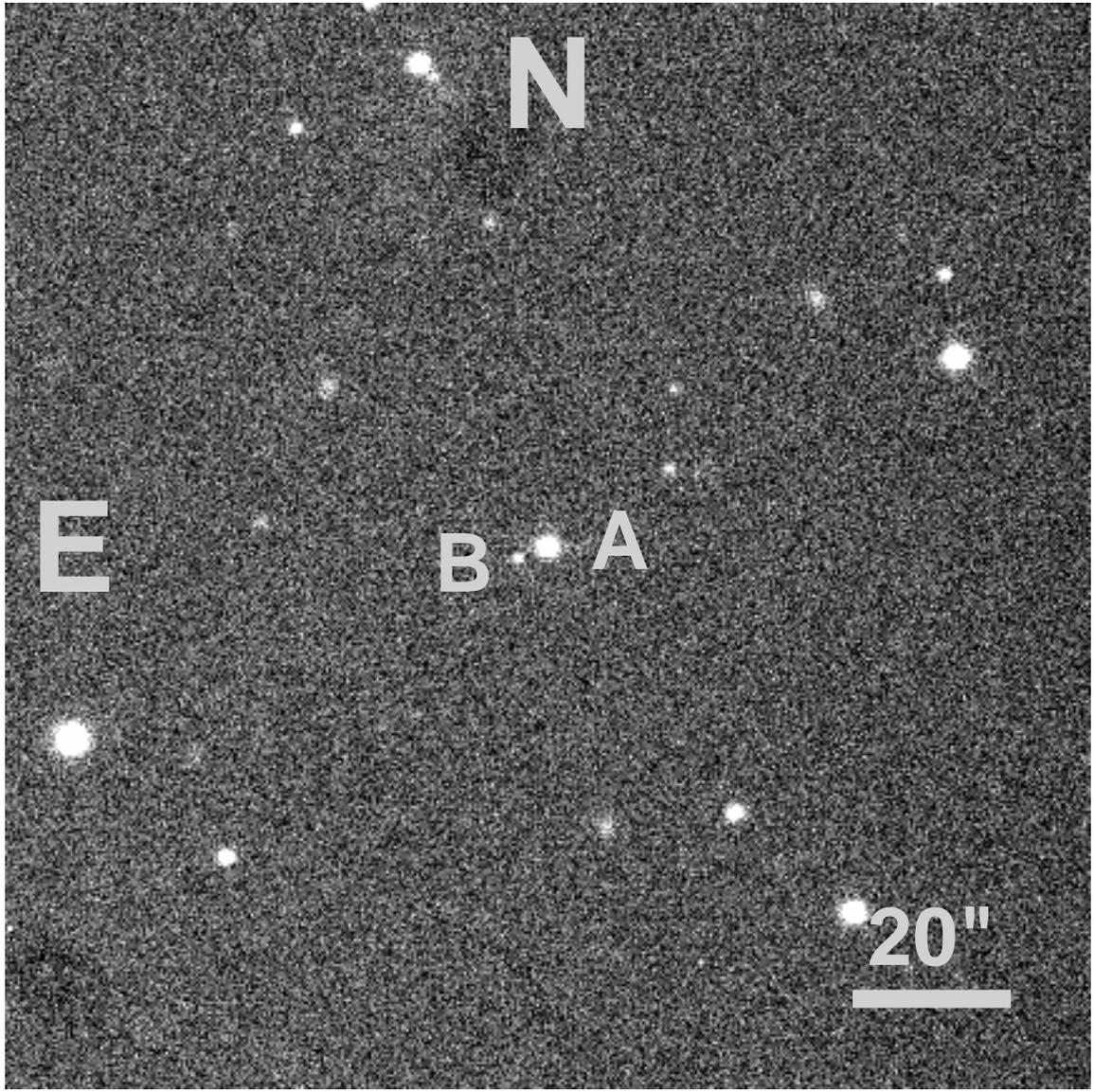
\end{document}